\documentclass[preprint,groupedaddress,floatfix,%
nofootinbib]{revtex4}

\usepackage{dcolumn}
\usepackage{graphicx}
\usepackage{epsfig}
\usepackage{hyperref}
\usepackage{graphicx}
\usepackage{dcolumn}
\usepackage{longtable}
\usepackage{amsmath,amssymb,bm}

\newcommand{\mathe}{\mathrm{e}}

\begin{document}

\title{A note on the architecture of spacetime geometry}
\author{Fen Zuo\footnote{Email: \textsf{18802711412@139.com}}}
\affiliation{School of Physics, Huazhong University of Science and Technology, Wuhan 430074, China}

\begin{abstract}
Recently the $\text{SU}(2)$ spin-network states in loop quantum gravity is generalized to those of the corresponding affine Lie algebra. We show that if one literally starts from the full $\text{SL}(2,\mathbb{C})$ group, this procedure naturally leads to the Bekenstein-Hawking formula of the entanglement entropy for any macroscopic spacetime region. This suggests that a smooth spacetime geometry could be recovered in such a way, as conjectured by Bianchi and Myers. Some comparison with Xiao-Gang Wen's string-net picture of gauge theory is made.
\end{abstract}
 \maketitle
\section{Motivation}
An interesting conjecture was recently made by Bianchi and Myers~\cite{Bianchi:2012ev}. The conjecture states that, for any sufficient large region in a smooth spacetime, the leading contribution to the entanglement entropy is given by the Bekenstein-Hawking formula~\cite{Bekenstein:1973ur,Hawking1974,Hawking:1974sw}. Various evidences for such a conjecture are collected. In particular, the spin network construction, in the loop quantum gravity~(LQG) approach~(for a review, see~\cite{Rovelli:2011eq}), is argued to satisfy such a requirement. However, a careful analysis shows that the spin-network states could at most provide a logarithmic contribution~\cite{Hamma:2015xla}. This indicates that there are some hidden degrees of freedom missing in the common LQG construction.

A nice generalization to include extra hidden degrees of freedom was suggested in~\cite{Ghosh:2014rra}, by extending the $\mathfrak{su}(2)$ algebra to an affine Lie algebra. Intuitively, one simply treats the spin-network punctures on the horizon as small circles, on which extra states could be excited. This is a common way to include quasiparticle excitations in a topological theory, see e.g., the related discussion in~\cite{Freedman2004428}. With such a generalization, one could therefore use the power techniques of $2d$ conformal field theory~(CFT)~(see~\cite{CFT} for original references and details) to make the state-counting on the horizon, or more generally on the entangling surface. While a CFT description of the $3d$ Ba\~{n}ados-Teitelboim-Zanelli~(BTZ) black hole~\cite{Banados:1992wn,Banados:1992gq} turns out to be very successful~\cite{Carlip:1994gy,Strominger:1997eq}, the state-counting here seems to involve some subtleties~\cite{Carlip:2014bfa}. A formal relation was derived in~\cite{Carlip:2015mea}~(see also~\cite{Pranzetti:2014tla}). In this short note we will try to clarify some of the subtleties, by adapting the CFT framework in~\cite{Carlip:1994gy} to the LQG approach.

\section{``Stringy'' spin network and entanglement entropy}
Essentially one is seeking an affine-Lie-algebra generalization of the usual $\text{SU}(2)$ spin-network states~\cite{Ghosh:2014rra}, in order to include the hidden degrees of freedom, and to take into account of a cosmological constant~\cite{Haggard:2014xoa,Haggard:2015yda,Haggard:2015kew}. In~\cite{Ghosh:2014rra} the full internal Lorentz group, or its cover group $\text{SL}(2,\mathbb{C})$, is first reduced to $\text{SU}(2)$ as usual. Then one extends the Lie algebra to the corresponding affine Lie algebra. Here we want to start literally from the $\text{SL}(2,\mathbb{C})$ group, as recently investigated in~\cite{Carlip:2015mea}. A systematic treatment and construction will be given in a future paper~\cite{Zuo:2017}. Instead we would like to make an intuitive comparison with the recent Witten-Costello construction for the integrable lattice models~\cite{Costello:2013sla,Witten:2016spx}, which gives a nice physical picture for the resulting algebraic structure.

Starting from the Chern-Simons theory, they introduce the loop extension (without central extension) of the gauge group. By properly recombining the loop parameter and one space coordinate, they obtain a mixed holomorphic/topological theory, which results the long-expected full-fledged Yang-Baxter equation with the spectral parameter~\cite{Yang:1967bm,Baxter:1972hz}. Due to the recombination, the original three dimensional symmetry in Chern-Simons theory is reduced to the 2-dimensional symmetry in the integrable lattice models~\cite{Costello:2013sla,Witten:2016spx}. Moreover, since the loop parameter and the space coordinate naturally have a holomorphic combination and an anti-holomorphic one, the full theory should contain two sectors. Similar considerations in the present case then lead us from the Lorentz group to the doubled affine Lie algebra $\widehat{\mathfrak{so}}(2,1)_k\otimes \overline{\widehat{\mathfrak{so}}(2,1)_k}$. This will be our starting point for the following discussion.

We write out the $\widehat{\mathfrak{so}}(2,1)_k\otimes \overline{\widehat{\mathfrak{so}}(2,1)_k}$ current algebra explicitly, following the notation in~\cite{Carlip:1994gy}. Some slight changes will be made in order to make comparison with string theory~\cite{GSW-1987}. The $\text{SO}(2,1)$ generators are chosen as
\begin{equation}
(T_a)_b^{~c}=-\epsilon _{abd}\eta^{dc},~\eta_{ab}=\mbox{diag}(-1,1,1),~~\epsilon_{012}=1,
\end{equation}
together with the structure constants and killing metric
\begin{equation}
f_{ab}^c=\epsilon_{abd}\eta^{dc},~~~\hat g_{ab}=2\eta_{ab}.
\end{equation}
The affine algebra could be expressed as
\begin{eqnarray}
\left[J^a_m,J_n^b\right]  & =  & \mathrm{i} f^{ab}_c J_{m+n}^c+km \hat g^{ab}\delta_{m+n,0},   \\
\left [\tilde J^a_m, \tilde J_n^b \right]  &  =  & \mathrm{i} f^{ab}_c \tilde J_{m+n}^c+km \hat g^{ab} \delta_{m+n,0}.
\end{eqnarray}
The $J^a_0$ and $\tilde J^a_0$ currents generate the corresponding global symmetry algebra. As shown in~\cite{Ghosh:2014rra}, the usual $\text{SU}(2)$ spin-network states could be extended to the affine case. In particular, the area operator, usually defined from the $\text{SU}(2)$ Casimir~\cite{Rovelli:1994ge}, could be directly generalized. An immediate question appears: how should we generalize the area operator to the present case? While the usual definition is kept in~\cite{Ghosh:2014rra}, we will try to argue that it is not proper from a local point of view. To see how a local definition could be achieved, we would like to take the large-$k$ limit instead.

The proper way of taking such a limit has been suggested in~\cite{Carlip:1994gy}. The relation between the asymptotic and the near horizon algebra of BTZ black holes has also been discussed recently in~\cite{Afshar:2016wfy,Afshar:2016uax}. One rescales the currents as
\begin{equation}
\alpha_m^a\equiv \sqrt{\frac{2}{k}}J_m^a,~~~~~~\tilde \alpha_m^a\equiv \sqrt{\frac{2}{k}}\tilde J_m^a.\label{eq.alpha}
\end{equation}
Then in the large-$k$ limit, we obtain an infinite set of decoupled $u(1)$ commutators
\begin{eqnarray}
\left[\alpha^a_m,\alpha_n^b\right]  & =  &   m  \eta^{ab}\delta_{m+n,0},   \\
\left[\alpha^a_m,\tilde \alpha_n^b\right]  & =  & 0, \\
\left[\tilde \alpha^a_m, \tilde \alpha_n^b\right]  & =  &   m  \eta^{ab}\delta_{m+n,0}.
\end{eqnarray}
Except for the identification $\alpha_0^a=\tilde \alpha_0^a$~\footnote{A detailed analyzes of this condition will be given in~\cite{Zuo:2017}. Here we simply ignore it, since it only affects the subleading term of the entropy.}, this is essentially the spectrum of a $(2+1)d$ closed string theory~\cite{GSW-1987}. However, the physical interpretations would be quite different. In string theory, these $u(1)$ currents are directly related to the target spacetime coordinates. Here in LQG, they are related to the Ashtekar connection~\cite{Ashtekar:1986yd}. Due to this, the physical observables are also different. In particular, the would-be string mass will be identified as a local area, up to normalization. Explicitly, we define the area as
\begin{equation}
A\equiv \sqrt{\frac{-\alpha_0^2-\tilde \alpha_0^2}{2}},\label{eq.area}
\end{equation}
where the contraction is made with $\eta_{ab}$. The area is supposed to be measured in Planck units with $l_p^2=8\pi \hbar G$, $G$ being the Newton constant. One could check that the above definition is in accordance with the re-scaling (\ref{eq.alpha}). In particular, the usual spin-network link with a finite $\text{SU}(2)$ spin $j$, after an internal Wick rotation, possesses a vanishing local area. In order to acquire a finite $A$, the original Casimirs $J_0^2$ and $\tilde J_0^2$ must be of order $k$ as $k\to \infty$. Notice that this double-scaling limit is different from the recent analyzes in~\cite{Haggard:2014xoa,Haggard:2015yda,Haggard:2015kew}. One may worry about the reality of the area definition. Later we will show that for a macroscopic surface the total area is always well defined.

The corresponding Virasoro generators are easily derived. In particular,
\begin{equation}
L_0=\frac{1}{2}\sum_{n=-\infty}^{\infty}\alpha_{-n}\cdot \alpha_n,\quad \tilde L_0=\frac{1}{2}\sum_{n=-\infty}^{\infty}\tilde \alpha_{-n}\cdot \tilde \alpha_n.
\end{equation}
One could now see that the definition (\ref{eq.area}) is directly related to the eigenvalues of $L_0$ and $\tilde L_0$ on highest-weight states~\cite{CFT,GSW-1987}. As in~\cite{Carlip:1994gy}, we ignore the normal ordering ambiguity which will be irrelevant for the state-counting. The construction of the Fock space is then essentially the same as string theory~\cite{GSW-1987}. One defines the vacuum as the state annihilated by all the generators $\alpha_n,~\tilde \alpha_n$ with positive $n$. The negative-$n$ generators are then treated as creating operators, which acting on vacuum create the excited states. These excited state could be labeled by the eigenvalues of the Virasoro operator $L_0$ and $\tilde L_0$, which are simply the sum of the zero mode contribution plus the excitation level~\cite{CFT,GSW-1987}:
\begin{equation}
L_0=\frac{\alpha_0^2}{2}+N,\quad \quad \tilde L_0=\frac{\tilde \alpha_0^2}{2}+\tilde N.
\end{equation}

Now another crucial difference from string theory appears. In string theory one imposes the Virasoro conditions on the physical states, which turns out to be very restrictive. One could think of the Virasoro conditions as a $2d$ version of the Hamiltonian and diffeomorphism constraints in the LQG approach~\cite{Carlip:1994gy,Rovelli:2015gwa}. However, the success in~\cite{Carlip:1994gy} suggests that we should instead impose only the global condition:
\begin{equation}
L_0=0,\quad \quad \tilde L_0=0. \label{eq.Virasoro}
\end{equation}
From this one immediate fixes the area of the excited state as
\begin{equation}
A=\sqrt{N+\tilde N}.
\end{equation}
Similar relation has also been found from string theory consideration~\cite{Susskind:1993ws,Halyo:1996vi,Halyo:1996xe},
through the calculation of the low energy absorption cross section~\cite{Halyo:1996xe}.

For a macroscopic surface with fixed $A$, the number of the allowed microscopic states could be easily derived~\cite{Huang:1970iq,GSW-1987,Carlip:1998qw,Carlip:2000nv}
\begin{equation}
n(A)\sim \mathe ^{2\pi \sqrt{6\cdot (N+\tilde N)/6} }/(N+\tilde N)^{(3+6)/4}\sim \mathe ^{2\pi A }/A^{9/2}.
\end{equation}
The corresponding entropy is simply
\begin{equation}
S \equiv \log n(A)\sim 2\pi A-{\mathcal O}(\log A),
\end{equation}
which is essentially the Bekenstein-Hawking formula~\cite{Bekenstein:1973ur,Hawking1974,Hawking:1974sw,Bianchi:2012ev}. As mentioned before, further relation between the zero modes $\alpha_0^a$ and $\tilde \alpha_0^a$ will only change the logarithmic term~\cite{Zuo:2017}.
Similar derivation of black hole entropy has been obtained in the early days from string theory~\cite{Susskind:1993ws,Halyo:1996vi,Halyo:1996xe}, together with some explanations of the specific central charge~\cite{Callan:1996dv,Tseytlin:1996qg}. Here we want to emphasize some important facts. First, the correct coefficient is only recovered with all the states taking into account. In other words, they may be ghost states with negative norms, which are usually eliminated by the Virasoro conditions together with specific choice of the spacetime dimension. This may not be a problem, since we are constructing the spacetime itself instead of a physical observable. Secondly, the full $\text{SL}(2,\mathbb {C})$ group has to be considered from the beginning, instead of the $\text{SU}(2)$ subgroup commonly studied. Finally, the above counting actually includes only the contributions of single-string states. Since only a single global constraint (\ref{eq.Virasoro}) is imposed, the whole system could always be treated effectively as a single string. An explicit example is shown in~\cite{Carlip:2015mea}. A full construction and interpretation will be given in~\cite{Zuo:2017}.

\section{Discussion and outlook}
We would like to make some comparison with the usual $\text{SU}(2)$ spin-network construction. As emphasized, a spin-$j$ link in the $\text{SU}(2)$ spin network possesses a vanishing local area as $k\to \infty$, according to the definition (\ref{eq.area}). Due to this, each link individually satisfies the constraint (\ref{eq.Virasoro}) at zero excitation level. By analogy with gauge theory~\cite{Levin:2004mi,Wen2004}, we may call such spin-network links ``confining'' states. In contrast, the global states counted in the previous section could be named ``deconfined'' states, or ``string-condensed'' states. In the former case, only separated short strings are present. While in the latter, a macroscopic surface is represented by a single long string. The entropy calculation indicates that only the deconfined phase, or string-condensed phase could possibly describe a smooth spacetime geometry, as conjectured by Bianchi and Myers. The whole phase diagram is thus very similar to the string-net picture of gauge theory~\cite{Wen2004,Levin:2005vf}.

However, to rigorously confirm such a statement one should explicitly construct the spacetime geometry from the boundary data of the affine Lie algebra. This could in principle be done following the common procedure in the LQG framework. In particular, this is supposed be done in the covariant formalism, the so-called spin foam approach~(see~\cite{Baez:1997zt,Baez:1999sr,Perez:2012wv} for details). As mentioned, some progresses in this direction have been recently made in~\cite{Haggard:2014xoa,Haggard:2015yda,Haggard:2015kew}. Yet there could be another way to explicitly construct the geometry, through the string-net approach based on the quantum deformation of $\text{SL}(2,\mathbb {C})$~\cite{Levin:2004mi,Wen2004}. These two approaches are then expected to be related through certain equivalence between the quantum group and the corresponding affine Lie algebra.

\section*{Acknowledgments}
The author would like to thank Ling-Yan Hung for inspiring discussions and explanations on loop quantum gravity and string-net condensation. He also benefits from the workshop ``Entanglement at Fudan 2015---where quantum information, condensed matter and gravity meet'', held at Fudan University, Shanghai, China. The hospitality of the organizers is acknowledged. This work was partially supported by the National Natural Science Foundation of China under Grant No. 11405065 and No. 11445001.

%
\end{document}